\documentstyle[aps]{revtex}

\def \be   {\begin{equation}}

\def \ee   {\end{equation}}

\def \l {\label}

\begin{document}

\input epsf


\baselineskip=25pt



\title{Discrete fields, general relativity, other possible implications and experimental evidences\footnote{This paper supersedes hep-th/0006250}}
\author{Manoelito M de Souza\footnote{Permanent address:Departamento de
F\'{\i}sica - Universidade Federal do Esp\'{\i}rito Santo\\29065.900 -Vit\'oria-ES-Brazil- E-mail: manoelit@cce.ufes.br}}
\address{Centro Brasileiro de Pesquisas F\'{\i}sicas- CBPF\\R. Dr. Xavier Sigaud 150,\\
22290-180 Rio de Janeiro -RJ - Brazil}
\date{\today}
\maketitle

\begin{abstract}

\noindent The physical meaning, the properties  and the consequences of a discrete scalar
 field are discussed; limits for the validity of a mathematical description of fundamental
  physics in terms of continuous fields are a natural outcome of discrete fields with discrete
   interactions.  The discrete scalar field is ultimately the gravitational field of general relativity,
    necessarily, and there is no place for any other fundamental scalar field, in this context.
     Part of the paper comprehends a more generic discussion about the nature, if continuous or discrete,
      of fundamental interactions. There is a critical point defined by the equivalence between the two descriptions. Discrepancies
        between them can be observed far away from this point as a  continuous-interaction is always stronger below it and weaker above
         it than a discrete one. It is possible that some discrete-field manifestations have already been observed
          in the flat rotation curves of galaxies and in the apparent anomalous acceleration of the Pioneer spacecrafts. The existence
           of a critical point is equivalent to
            the introduction of an effective-acceleration scale which may put Milgrom's MOND on a more solid physical basis.
Contact is also made, on passing, with  inflation in cosmological theories and with Tsallis  generalized one-parameter statistics which is regarded as 
 proper for discrete-interaction systems. The validity of Botzmann statistics is then reduced to idealized asymptotic states which, rigorously,
  are reachable only after an infinite number of internal interactions . Tsallis parameter is then a measure of how close a system
  is from its idealized asymptotic state.

\end{abstract}

\begin{center}


\end{center}
\section{Introduction}
Although it is considered that a scalar field has not been observed in
nature as a fundamental field its use as such is very frequent in the
modern literature, particularly in elementary particles, field theory and
cosmology. Here we will apply to the scalar field the concepts and results
developed in the reference \cite{hep-th/0006237}, referred here as the
paper I, where the concept of a discrete field was introduced and its wave
equation and its Green's function discussed. The standard field and its
formalism, which for a distinction, we always append the qualification
continuous, are retrieved from an integration over the discrete-field
parameters. Remarkable in the discrete field is that it has none of the
problems that plague the continuous one so that the meaning and origin of
these problems can be left exposed on the passage from the discrete to the
continuous formalism \cite{hep-th/0006237}. Although the motivations for
the introduction of a generic discrete field in paper I have being made on
pure physical grounds of causality, a deeper discussion about its physical
interpretation have been left for subsequent papers on specific fields.
This discussion will be retaken here with the simplest structure of a
field, the scalar one. It would be a too easy posture to see the discrete
field as just an ancillary mathematical construct devoid of any physical
meaning, a vision that could be re-enforced with the discrete  field as a
pointlike signal. The idea of a pointlike field may sound weird at a first
sight but this represents the same symmetry of quantum field theory where
fields and sources are equally treated as quantized fields. Here they are
seen from a reversed classical perspective. Besides, pointlike object is
not a novelty in physics and one of the major motivations of the string
theory is of avoiding \cite{Polchinski} infinities and acausalities in the
fields produced by point sources; problems that do not exist for the
discrete field, according to the reference \cite{hep-th/9610028}.

This paper is structured in the following way. Section II, on the sake of a
brief review of the mathematical definition of discrete fields, is a recipe
on how to pass from a continuous to a discrete field formalism, and
vice-versa. The discrete scalar field, its wave equation, its Lagrangian
and its energy tensor are discussed in Section III. The paper major
contribution begins in Section IV that discusses the consequences of
discrete interactions for the mathematical description of the physical
world. Then it gains generality as the discussions leaves the specificity
of scalar interactions widening to the universality of all fundamental
interactions.  Calculus (integration and differentiation) which is based on
the opposite idea of smoothness and continuity, has its full validity for
describing dynamics restricted then to a very efficient approximation  in
the case of a high density of interaction points, such that the concept of
acceleration as a continuous change of velocity may be introduced in an
effective physical description of fundamental interactions. This seems to
be an answer to the Wigner's pondering \cite{Wigner} about the reasons
behind the unexpected effectiveness of mathematics on the physical
description of the world. It is argued in Section V, after the results of
the Section IV, that the scalar field must necessarily describe the
gravitational interaction of general relativity whose character of a
second-rank tensor is assured by the way the scalar field is attached to
the definition of the metric tensor. After decoding the physical meaning of
the scalar-field sources one is led to the unavoidable conclusion that
there is no place, in this context, for the existence of any other
fundamental scalar field. This has deeper theoretical and observational
implications, discussed in Section VI, where the possibility that
consequences of discrete gravity have already been observed is considered.
This would set experimental limits on the validity of general relativity as
an effective  field theory. Contact is made, on passing, with inflationary
cosmology and with the Tsallis's statistics. The paper ends with some
concluding remarks in Section VII.
\section{From continuous to discrete}

For a concise introduction of the discrete-field concept it is convenient to replace the Minkowski spacetime flat geometry by a conical projective one in an embedding (3+2) flat spacetime:
\be
\l{def}
\{x\in R^4\}\Rightarrow\{x,x^5\in R^5{\big|}(x^5)^2+x^2=0\},
\ee
where $x\equiv({\vec x},t)$ and $x^2\equiv\eta_{\mu\nu}x^{\mu}x^{\nu}=|{\vec x}|^2-t^2.$ So a change $\Delta x^5$ on the fifth coordinate, allowed by the constraint $(\Delta x^5)^2+(\Delta x)^2=0,$   is a Lorentz scalar that can be interpreted as a change $\Delta\tau$ on the proper-time  of a physical object propagating across an interval $\Delta x:\quad \Delta x^5=\Delta\tau=\pm\sqrt{(\Delta t)^2-(\Delta{\vec x})^2}.$

The constraint
\be
\l{hcone}
(\tau-\tau_{0})^2+(x-x_{0})^2=0
\ee
defines a double hypercone with vertex at $(x_{0},\tau_{0}),$ whilst
\be
\l{hplane}
(\tau-\tau_{0})+f_{\mu}(x-x_{0})^{\mu}=0
\ee
defines a family of hyperplanes tangent to the double hypercone and labelled by their normal\footnote{The Eq. (\ref{hplane}) can be written in $R^5$ as 
$f_{M}\Delta x^{M}=0,\;\;M=1,2,3,4,5$ with $f_{M}=(f_{\mu},1)$} $f_{\mu}$, a constant four-vector. The intersection of the double hypercone with a hyperplane 
defines its $f$-generator tangent to $f^{\mu}$  ($f^{\mu}:=\eta^{\mu\nu}f_{\nu}).$ A discrete field is a field defined with support on this intersection 
(extended causality) in 
contraposition  \cite{hep-th/0006237} to the continuous field, defined with support on a hypercone (local causality):
\be
\l{df}
\phi_{f}(x,\tau):=\phi(x,\tau){\Big|}_{{\Delta\tau+f.\Delta x=0}\atop{\Delta\tau^2+\Delta x^2=0}}:=\phi{\Big|}_{f}.
\ee
The symbol ${\big|}_{f}$ is a short notation for the double constraint in the middle term of Eq. (\ref{df}). The constraint (\ref{df}) induces the 
directional derivative (along the fibre $f$, the hypercone $f$-generator)
\be
\l{dd}
\nabla_{\mu}\phi_{f}(x,\tau):=(\partial_{\mu}-f_{\mu}\partial_{\tau})\phi_{f}(x,\tau).
\ee

An action for a discrete scalar field is
\be
\l{da1}
S_{f}=\int d^5x{\Big\{}\frac{1}{2}\eta^{\mu\nu}\nabla_{\mu}\phi_{f}(x,\tau)\nabla_{\nu}\phi_{f}(x,\tau)-\phi_{f}(x,\tau)\rho(x,\tau){\Big\}},
\ee
where $d^5x=d^4xd\tau$, and $\rho(x,\tau)$ is the source for the scalar field. There can be no mass term in a discrete-field Lagrangian because it would 
imply on a hidden breaking of the Lorentz symmetry with non-propagating discrete solutions of the  field equations. In other words no physical object could 
be described by such a Lagrangian with an explicit mass term. Nevertheless, as discussed  in paper I, the action (\ref{da1}) still describes both, massive 
and massless fields. The mass of a massive discrete field is implicit on its propagation with a non-constant proper-time. Eq. (\ref{da1}) is a scale-free 
action expressing the (1+1)-dynamics of a discrete field, massive or not, on a fibre $f$; a mass term would break its conformal symmetry 
\cite{hep-th/0006237}.

Then the field equation and the tensor energy for a discrete field are, respectively,
\be
\l{dfe}
\eta^{\mu\nu}\nabla_{\mu}\nabla_{\nu}\phi_{f}(x,\tau)=\rho(x,\tau),
\ee
\be
\l{det}
T^{\mu\nu}_{f}=\nabla^{\mu}\phi_{f}\nabla^{\nu}\phi_{f}-\frac{1}{2}\eta^{\mu\nu}\nabla^{\alpha}\phi_{f}\nabla_{\alpha}\phi_{f}.
\ee
They must be compared to the standard expressions for the continuous field:
\be
\l{usfe}
(\eta^{\mu\nu}\partial_{\mu}\partial_{\nu}-m^2)\phi(x)=\rho(x),
\ee
\be
\l{ts}
T^{\mu\nu}(x)=\partial^{\mu}\phi\partial^{\nu}\phi-\frac{1}{2}\eta^{\mu\nu}\partial^{\alpha}\phi\partial_{\alpha}\phi
\ee
which can be obtained from the action
\be
\l{a}
S=\int d^{4}x{\Big\{}\frac{1}{2}\eta^{\mu\nu}\partial_{\mu}\phi_{f}\partial_{\nu}\phi_{f}-\frac{m^2}{2}\phi^2-\phi(x)\rho(x){\Big\}},
\ee
So, the passage from a continuous to a discrete field formalism can be summarized in the following schematic recipe (the arrows indicate replacements):
\be
\l{recipe}
\cases{\{x\}\Rightarrow\{x,x^5\};\cr
\phi(x)\Longrightarrow \phi(x,\tau){\Big |}_{f};\cr
\partial_{\mu}\Rightarrow\nabla_{\mu},\cr}
\ee
accompanied by a dropping of the mass term from the Lagrangian. Moreover a discrete field requires a discrete source \cite{hep-th/0006237}. A continuous $\rho(x)$ is replaced by a discrete set of pointlike sources $\rho(x,\tau)$. Any apparent continuity is reduced to a question of scale in the observation. $\rho(x,\tau)$ is, like  $\phi_{f}(x,\tau)$, a discrete field defined on a hypercone generator too, which just for simplicity, is  not being considered here. This is a symmetry between fields and sources: they are all discrete fields, and the current density of one is the source of the other.

Reversely, in the passage from discrete to continuous, the continuous field and its field equations are recuperated in terms of effective average fields smeared over the hypercone
\be
\label{s}
\Phi(x,\tau)=\frac{1}{2\pi}\int d^{4}f\delta(f^2)\Phi_{f}(x,\tau).
\ee
This passage provokes the appearing of the mass term and the breaking of the conformal symmetry of the action (\ref{da1}).
This has been explicitly proved, for both the massive and the massless fields, in the reference \cite{hep-th/0006237}.

\section{The discrete scalar field}

Comparing the actions of Eqs. (\ref{da1}) and (\ref{a}) one should observe that the first one contains explicit manifestations only of the constraint (\ref{hplane}) through the use of the directional derivatives (\ref{dd}), but not of the constraint (\ref{hcone}). This one is only dynamically introduced through the solutions of the field equation, like it happens also (local causality) in the standard formalism of continuous fields \cite{Jackson}. As a matter of fact all the information contained in the new action (\ref{da1}) can be incorporated in the old action (\ref{a}), without its mass term, with the simple inclusion of the constraint (\ref{hplane})
\be
\l{ar}
S_{P}=\int d^{4}xd\tau\delta(\Delta\tau+f.\Delta x){\Big\{}\frac{1}{2}\eta^{\mu\nu}\partial_{\mu}\phi\partial_{\nu}\phi-\phi(x,\tau)\rho(x,\tau){\Big\}},
\ee
as the very  restriction to the hyperplane (\ref{hplane}) by itself implies on the whole recipe (\ref{recipe}). $P$ in Eq. f(\ref{ar}) stands for any 
generic fixed point, the local hypercone vertex: $P=(x_{0},\tau_{0}),$ $\Delta\tau=\tau-\tau_{0}$ and $\Delta x=x-x_{0}.$ Local causality, dynamically 
implemented through the field equations, imply that the field propagates on a hypercone (the lightcone, if a massless field) with vertex on $P,$  which is 
an event on the world  line of $\rho(x,\tau)$. The constraint (\ref{hplane}) included in this action (\ref{ar}) further restricts the field to the fiber 
$f$, expressing an extended concept of causality \cite{hep-th/0006237,hep-th/0006214}.

Whereas there is no restriction on $\rho(x)$ for a continuous field, for a discrete one, as already mentioned, it must be a discrete set of point sources. A continuously extended source would not be consistent as it would produce a continuous field. The source of a discrete scalar field is given by
\be
\l{rof}
\rho(x,t_{x}=t_{z})=q(\tau_{z})\delta^{(3)}({\vec x}-{\vec z}(\tau_{z}))\delta(\tau_{x}-\tau_{z}),
\ee
where $z(\tau)$ is its world  line parameterized by its proper time $\tau$; $q(\tau)$ is the scalar charge whose physical meaning will be made clear later. 
The sub-indices in $t$ and $\tau$ specify the respective events $x,\;y$ and $z$. That $t_{x}$ must be equal to $t_{z}$ on the left-hand side  of Eq. 
(\ref{rof}) is a consequence of the deltas on its right-hand side and of the constraint (\ref{hcone}). Initially, it is assumed that both 
${\dot q}\equiv\frac{dq}{d\tau}$ and ${\ddot q}\equiv\frac{d{\dot q}}{d\tau}$ exist and that they may be
non null.
The field eq. (\ref{dfe}) is solved by
\be
\l{phif}
\phi_{f}(x,\tau)=\int d^5yG_{f}(x-y,\tau_{x}-\tau_{y})\rho(y,\tau_{y})
\ee
with
\be
\l{boxGf}
\eta^{\mu\nu}\nabla_{\mu}\nabla_{\nu}G(x,\tau)=\delta^{(5)}(x)=\delta(\tau)\delta^{(4)}(x).
\ee
The discrete Green's function associated to the Klein-Gordon operator is given\cite{hep-th/0006237} by
\be
\l{pr9}
G_{f}(x,\tau)=\frac{1}{2}\theta(bf^4t)\theta(b\tau)\delta(\tau+ f.x),\;\;\;\;{\vec x}_{{\hbox{\tiny T}}}=0,
\ee

where $b =\pm1,$ and $\theta (x)$ is the Heaviside function, $\theta(x\ge0)=1$ and $\theta(x<0)=0.$ The labels {\tiny L} and {\tiny T} are used as an 
indication of, respectively,  longitudinal and transversal with respect to the space part of $f$: ${\vec f}.{\vec x}_{\hbox{\tiny T}}=0$ and 
$ x_{{\hbox{\tiny L}}}=\frac{{\vec f}.{\vec x}}{|{\vec f}|}$.

Remarkably $G_{f}(x,\tau)$ does not depend on anything outside its support, the fibre $f$, as stressed by the append ${\vec x}_{{\hbox{\tiny T}}}=0$. One 
could retroactively use this knowledge in the action (\ref{da1}) for rewriting it as
\be
\l{sf3}
S_{f}=\int d^{5}x\delta^{(2)}({\vec x}_{\hbox{\tiny T}}){\Big\{}{1\over2}\eta^{\mu\nu}\nabla_{\mu}\phi_{f}\nabla_{\nu}\phi_{f}-\phi_{f}(x,\tau)\rho(x,\tau)
{\Big\}},
\ee
just for underlining that the fibre $f$ induces a conformally invariant (1+1) theory of massive and massless fields, embedded in a (3+1) theory, as 
generically discussed in paper I. Actually, the factor $\delta^{(2)}({\vec x}_{\hbox{\tiny T}})$ is an output of the actions (\ref{da1}) or (\ref{ar}) 
(it is not necessary to put it in there by hand) and it can never be incorporated as a factor in the definition (\ref{pr9}) of $G_{f}(x,\tau)$, except 
under an integration sign as in Eqs.
(\ref{phif}) and (\ref{sf3}).

Then one could, just formally, use
\be
\l{dsf}
\rho_{[f]}(x-z,\tau_{x}-\tau_{z})=q(\tau)\delta(\tau_{x}-\tau_{z})\delta(t_{x}-t_{z})\delta(x_{\hbox{\tiny L}}-z_{\hbox{\tiny L}}),
\ee
where $\rho_{[f]}$ represents

the source density $\rho$ stripped of its explicit ${\vec x}_{\hbox {\tiny T}}$-dependence, for reducing the action to
\be
\l{21}
S_{f}=\int d\tau_{x}dt_{x}dx_{\hbox{\tiny L}}{\Big\{}{1\over2}\eta^{\mu\nu}\nabla_{\mu}\phi_{f}\nabla_{\nu}\phi_{f}-\phi_{f}(x,\tau)\rho_{[f]}(x,\tau)
{\Big\}},
\ee
by just omitting the irrelevant transversal coordinates. Eq. (\ref{da1}) then, after its output Eq. (\ref{pr9}), is formally equivalent to Eq. (\ref{21}).
 But we should observe that this is no more than a formal expression once $\rho_{[f]}$ then represents just an event, the intersection of the worldline of 
 $\rho(x)$, whose support is not $f$, with the fibre $f$, support of $\phi_{f}(x)$. See the Figure 1.
\vglue13cm
\hglue-1.0cm
\begin{minipage}[]{7.0cm}
\parbox[t]{5.0cm}{
\begin{figure}
\vglue-13cm
\epsfxsize=200pt
\epsfbox{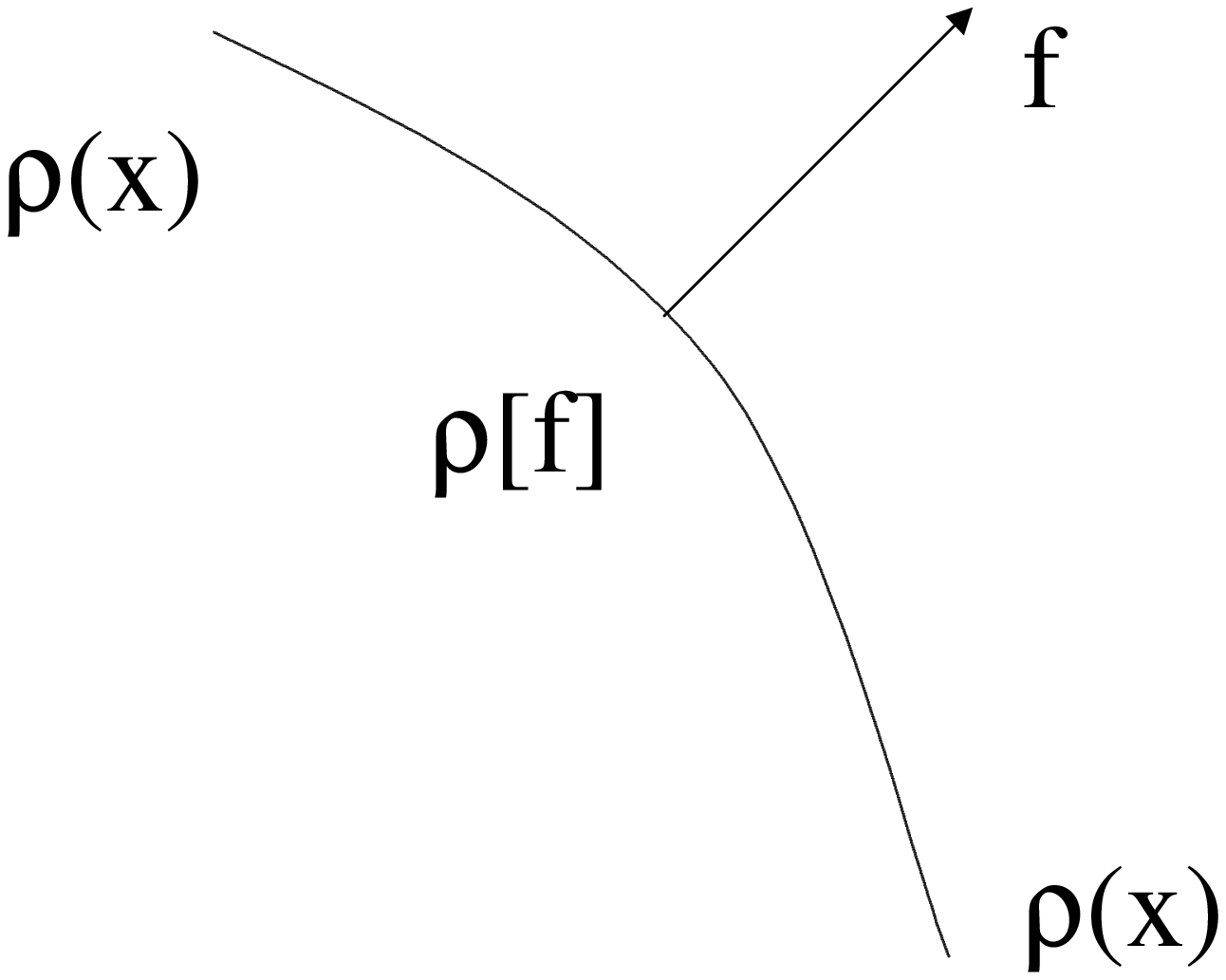}
\end{figure}}
\end{minipage}\hfill
\vglue-7cm
\hglue-1.0cm
\begin{minipage}[]{7.0cm}
\begin{figure}
\hglue9.0cm
\parbox[t]{5.0cm}{\vglue-6cm
\caption{The meaning of $\rho_{[f]}$: the value of $\rho(x)$ at the specific point defined by the intersection of the worldline of $\rho(x)$, whose 
support is not $f$, with the fibre $f$, support of $\phi_{f}(x)$. }}
\end{figure}
\end{minipage}

\vglue-1cm

The solutions from Eq. (\ref{usfe}), with $m=0$, for a point source are well known massless spherical waves propagating 
(forwards or backwards in time) on a lightcone in contradistinction to the solutions (\ref{pr9}) that are, massive or massless point signals 
propagating always forwards in time on a straight line, a generator of the hypercone (\ref{hcone}). Being massive or massless is determined by 
$\tau$ being constant or not, as discussed in paper I. For a massive field, its mass and its timelike four velocity are hidden behind a lightlike 
$f$ and a non-constant $\tau$; they  become explicit only after the passage from discrete to continuous fields. But as it will be made clear in 
Section V, there is no point on considering a massive discrete scalar field because any discrete scalar field must be associated to the gravitational 
field of general relativity. So massive discrete scalar fields will not be considered here any further.
With $b=+1$ and $f^4\ge1$ which implies an  emitted field, one has from Eqs. (\ref{pr9}) and (\ref{rof}) that
$$
\phi_{f}(x,\tau_{x})=\int d^{5}y \theta(t_{x}-t_{y})\theta(\tau_{x}-\tau_{y})\delta[\tau_{x}-\tau_{y}+ f.(x-y)]q(\tau_{z})\delta^{4}(y-z)=$$
\be
=\int d\tau_{y} \theta(t_{x}-t_{y})\theta(\tau_{x}-\tau_{y})\delta[\tau_{x}-\tau_{y}+ f.(x-y)]q(\tau_{z}),
\ee
where an extra factor 2 accounts for a change of normalization with respect to Eq. (\ref{pr9}) due to the exclusion of the annihilated field (which corresponds in Eq. (\ref{s}) to the integration over the future lightcone). Then,
\be
\l{Af0}
\phi_{f}(x_{{\hbox{\tiny L}}},{\vec x}_{{\hbox{\tiny T}}}={\vec z}_{{\hbox{\tiny T}}},t_{x},\tau_{x}=\tau_{z})= \theta(t_{x}-t_{z})\theta(\tau_{x}-\tau_{z})q(\tau_{z}){\Big |}_{f.(x-z)=0}
\ee
or for short, just
\be
\l{Af1}
\phi_{f}(x,\tau)= q(\tau)\theta(t)\theta(\tau){\Big |}_{f}.
\ee

$\nabla\theta(t)$ and $\nabla\theta(\tau)$ do not contribute \cite{hep-th/0006237} to $\nabla \phi_{f},$ except at $x=z(\tau),$
as a further consequence of the field constraints. So, for $t>0$ and (therefore) $\tau\geq0$  one can write just
\be
\l{AfV}
\phi_{f}(x,\tau_{x})=q(\tau_{z}){\Big |}_{f}
\ee
\be
\l{dAf}
\nabla_{\nu}\phi_{f}=-f_{\nu}{\dot q}{\Big |}_{f}
\ee
With Eq. (\ref{dAf}) in Eq. (\ref{det}) one has
\be
\l{tsf1}
T^{\mu\nu}_{f}(x,\tau_{x})=f^{\mu}f^{\nu}{\dot q}^{2}{\Big |}_{f}
\ee
The field four-momentum, given by $\int T^{\mu\nu}n_{\nu}d\sigma$ for a continuous field, is reduced, thanks to the field pointlike character and to its independence from the transversal coordinates, to
\be
\l{psf}
p^{\mu}_{f}=T^{\mu\nu}_{f}n_{\nu}=f^{\mu}{\dot q}^{2}{\Big |}_{f}
\ee
where $n$ is a spacelike four vector \cite{hep-th/0006237} such that $n.f=1$. The conservation of the energy-momentum content of $\phi_{f}$ is assured then just by $f$ being lightlike, $f^2=0,$
\be
\l{emc}
\nabla_{\mu}T^{\mu\nu}_{f}=-2f_{\mu}f^{\mu}f^{\nu}{\dot q}{\ddot
q}{\Big |}_{f}=0. \ee
 It is justified naming $\phi_{f}$ a discrete field
because although being a field it is not null at just one space point at a
time; but it is not a distribution, a Dirac delta function, as it is
everywhere and always finite. Its differentiability, in the sense of having
space and time derivatives, is however assured by its dependence on $\tau$,
a known continuous spacetime function. It is indeed a new concept of field,
a very peculiar one, discrete and differentiable; it is just a finite
pointlike spacetime deformation projected on a null direction, with a well
defined and everywhere conserved energy-momentum. It is this discreteness
in a field that allows the union of wave-like and particle-like properties
in a same physical object (wave-particle duality); besides this implies
\cite{hep-th/9911233} finiteness and no spurious degree of freedom
(uniqueness of solutions).

 \section{Discrete physics}

According to Eq. (\ref{AfV}), the field $\phi_{f}$ is given, essentially,
by the charge at its retarded time, i.e. the amount of scalar charge at
$z$, the event of its creation.  It has a physical meaning, in the sense of
having an energy-momentum content, when and only when ${\dot q}\not=0$. So,
the emission or the absorption of a scalar field is, respectively,
consequence or cause of a change in the amount of scalar charge on its
source. This is so because emitting or absorbing a scalar field requires a
change in the state of its source which is so poor of structure that has
nothing else to change but itself, and this is fundamental for determining
the scalar-charge nature. The picture becomes clearer after recalling that
we are dealing with discrete field and discrete interactions  which implies
that the change in the state of a field source occurs at isolated events.
$q(\tau)$ is not a continuous function:
\be
\l{dV}
q(\tau):=\sum_{i}q_{\tau_{i+1}}{\bar\theta}(\tau_{i+1}-\tau){\bar\theta}(\tau-\tau_{i}),
\ee
where
\be
\l{dth}
{\bar\theta}(x)=\cases{1, &if $x>0$;\cr 1/2,& if $x=0$;\cr 0,& if $x<0$,\cr}
\ee
and the index $i$ labels the interaction points on the source worldline, $i=1,2,3\dots$. For a given $\tau$ only one, or at most two terms contribute to the sum in Eq. (\ref{dV})
\be
q(\tau)=\cases{q_{\tau_{j}}, & if $\tau_{j}<\tau<\tau_{j+1}$;\cr
        &\cr
        \frac{q_{\tau_{j-1}}+q_{\tau_{j}}}{2}, & if $\tau=\tau_{j}$;\cr
        &\cr
        q_{\tau_{j-1}}, & if $\tau_{j-1}<\tau<\tau_{j}$,\cr}
\ee
as indicated in  the graph of the Figure 2.
\vglue13cm
\hglue-2.0cm
\begin{minipage}[]{7.0cm}
\parbox[t]{5.0cm}{
\begin{figure}
\vglue-14cm
\epsfxsize=300pt
\epsfbox{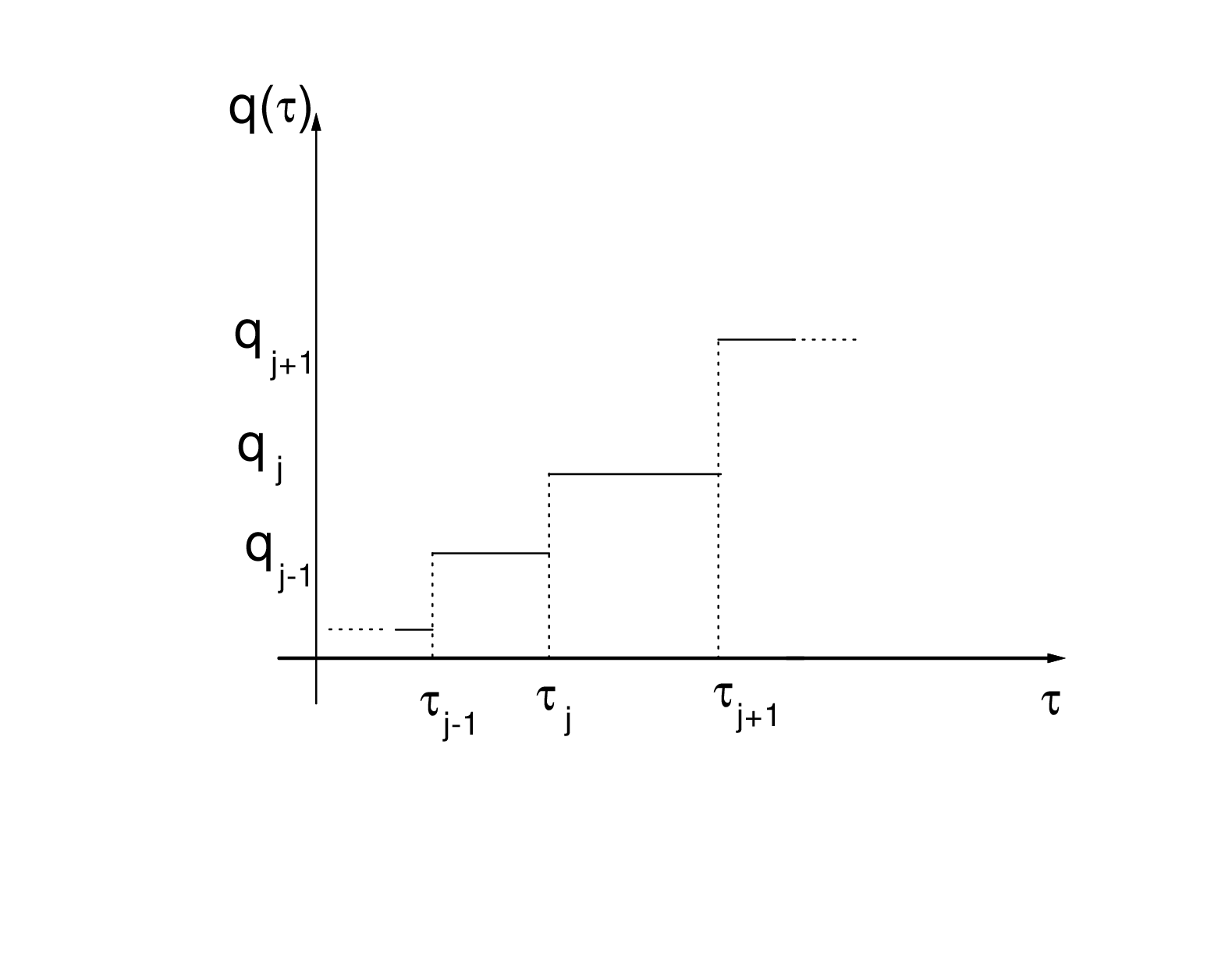}
\end{figure}}
\end{minipage}\hfill
\vglue-7cm
\hglue-1.0cm
\begin{minipage}[]{7.0cm}
\begin{figure}
\hglue9.0cm
\l{f3a}
\parbox[t]{5.0cm}{\vglue-6cm
\caption{Discrete changes on a discrete scalar charge along its worldline. A discrete scalar charge is so poor of structure that there is nothing else to 
change but itself. There is change in the state of a scalar source only at the interaction points on its worldline which is labelled by its proper time. 
If only the (discrete) interaction points are relevant the proper time may be treated as a discrete variable. In the limit of a worldline densely packed 
of interaction points a continuous graph is a good approximation.}}
\end{figure}
\end{minipage}

\vglue-1cm

The change in the state of the scalar source is not null only at the (discrete) interaction points and so, rigourously, it cannot  be defined as a time derivative, as there is no continuous variation, just a sudden finite change. The naive use of
\be
\l{dasdd}
{\dot q}=q(\tau)\delta(\tau-\tau_{z}),
\ee
would be just an insistence on an unappropriate continuous formalism, besides artificially introducing infinities where there is none. It means that one must replace time derivatives by finite differences
\be
\l{finitedif}
{\dot q}(\tau)\Rightarrow \cases{\Delta q_{\tau_{j}} & if $\tau=\tau_{j}$;\cr
                0 & if $\tau\neq\tau_{j}$,\cr}
\ee and a proper-time integration by a sum over the interaction points on
the charge. The existence and meaning of any physical property that
corresponds to a time derivative must be reconsidered at this fundamental
level. Velocity ($v)$ exists as a piecewise smoothly continuous function
(discontinuous at the interaction points). Acceleration ($a$) and
derivative concepts like force ($F$), power ($P$), etc rigorously do not
exist. We must deal with finite differences, respectively, the  sudden
changes of velocity ($v$), momentum ($p$) and energy ($E$):
\be
\l{reciped}
\cases{a\Rightarrow\Delta v\cr
F\Longrightarrow\Delta p \cr
P\Rightarrow\Delta E\cr}
\ee                                                                                                                                                                                                                                                                   
The observability of an interaction discreteness is in fact controlled by the ratio ($\frac{\Delta q_{j}}{\Delta\tau_{j}}$) of the two parameters 
$\Delta q_{j}$ and $\Delta\tau_{j}$ shown in the Figure 2,  as the validity of an approximative continuous description of fundamental interactions 
requires the
existence of 
\be
\l{ratio} {\dot q}_{j}= \frac{\Delta q_{j}\rightarrow0}{\Delta
\tau_{j}\rightarrow0}\neq0, \ee which is interpreted as a time derivative
of $q(\tau),$ taken as a smooth continuous function of $\tau.$ But actually
\be
\l{X}
\Delta X_{j}\rightarrow0
\ee
has the meaning that both discrete changes, $\Delta q_{j}$ and $\Delta \tau_{j}$, are
 smaller than their respective experimental thresholds of
 detectability, which,
  of course, is existing-technology dependent. 
  For two-body interactions $\Delta \tau_{j}$ is twice the flying time between them and
is then proportional to their space separation, $$\Delta\tau_{j}=\frac{2R}{c}.$$ See the Figure 3. For a large number of interacting bodies $\Delta \tau_{j}$ is a statistical 
average time-interval between two consecutive interaction events on one body worldline. 
 It decreases with the number of participants, and therefore, in the case of gravitational interaction, with the masses of the macroscopic interacting bodies.
\vglue13cm
\hglue-1.0cm
\begin{minipage}[]{7.0cm}
\parbox[t]{5.0cm}{
\begin{figure}
\vglue-13cm
\epsfxsize=250pt
\epsfbox{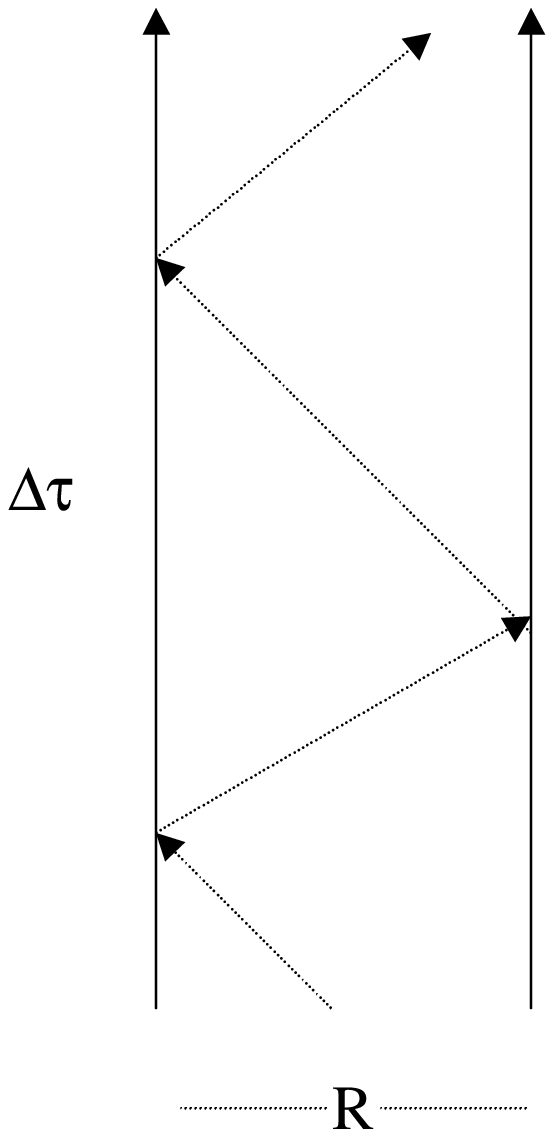}
\end{figure}}
\end{minipage}\hfill
\vglue-5cm
\hglue-1.0cm
\begin{minipage}[]{7.0cm}
\begin{figure}
\hglue7.0cm
\parbox[t]{5.0cm}{\vglue-6cm
\caption{Discrete two-body interactions. $\Delta\tau_{j}$ is the time interval between two consecutive interaction events on a worldline.}}
\end{figure}
\end{minipage}

\vglue-1cm

\vglue-1cm

$\Delta q_{j}$ is interaction dependent. It defines the interaction symmetry. 
Therefore,
$\frac{\Delta q_{j}}{\Delta\tau_{j}}$ would diverge if  $\Delta \tau_{j},$ but not
$\Delta q_{j}$,
would satisfy Eq. (\ref{X}), and it would unduly\footnote{Because the actual
interaction is not null.} be null in the case of only  $\Delta q_{j},$
but not $\Delta \tau_{j},$ satisfying it. The interaction strenght is described by the limit of $\frac{\Delta q_{j}}{\Delta\tau_{j}}$  (as a time 
derivative of $q(\tau)$) for a continuous interaction, and by both independent parameters $\Delta q_{j}$ and $\Delta \tau_{j}$ for a discrete one.  
For a discrete interaction the ratio  
$\frac{\Delta q_{j}}{\Delta\tau_{j}}$ has no special meaning.
 Both results, infinity and an undue zero, evince the existence of two demarcating points, a near and a far one, signalizing the inadequacy of the 
 approximative 
 continuous-interaction description. The two  points
 delimit the range of the ratio-parameter $\frac{\Delta q_{j}}{\Delta\tau_{j}}$ where there is no observationally detectable difference between a discrete
 and a continuous interaction. This defines the domain of validity of a continuous field as an effective physical description.  A continuous field is then
 stronger below the near point and weaker above the far one than its corresponding discrete field. Outside the range delimited by these points a
 discrete-interaction description must be used. This is schematically represented in Figure 4 that superposes, with two graphs $q\times R$, both the
 continuous and the discrete descriptions of a given interaction. For the sake of simplicity, the discrete description is also represented by a smooth and
 continuous curve. The region delimited by the two curves and the demarcating points is, by definition, not resolved with the present technology. The two
 demarcating points, near and far, represent the experimental resolution thresholds of the two descriptions. They are, by
 definition, dependent of the existing technology but there is, inside this region, a
 critical point of absolute equality of the two descriptions, defined by
 the two-curve crossing,  which is technology independent. The existence of this
 critical point sets a scale for the interaction strenght in terms of an effective
 time derivative of $q(\tau)$. The discrete field formalism, we remind, being
 conformally symmetric \cite{hep-th/0006237}, is scale free.
\vglue13cm
\hglue-1.0cm
\begin{minipage}[]{7.0cm}
\parbox[t]{5.0cm}{
\begin{figure}
\vglue-13cm
\epsfxsize=200pt
\epsfbox{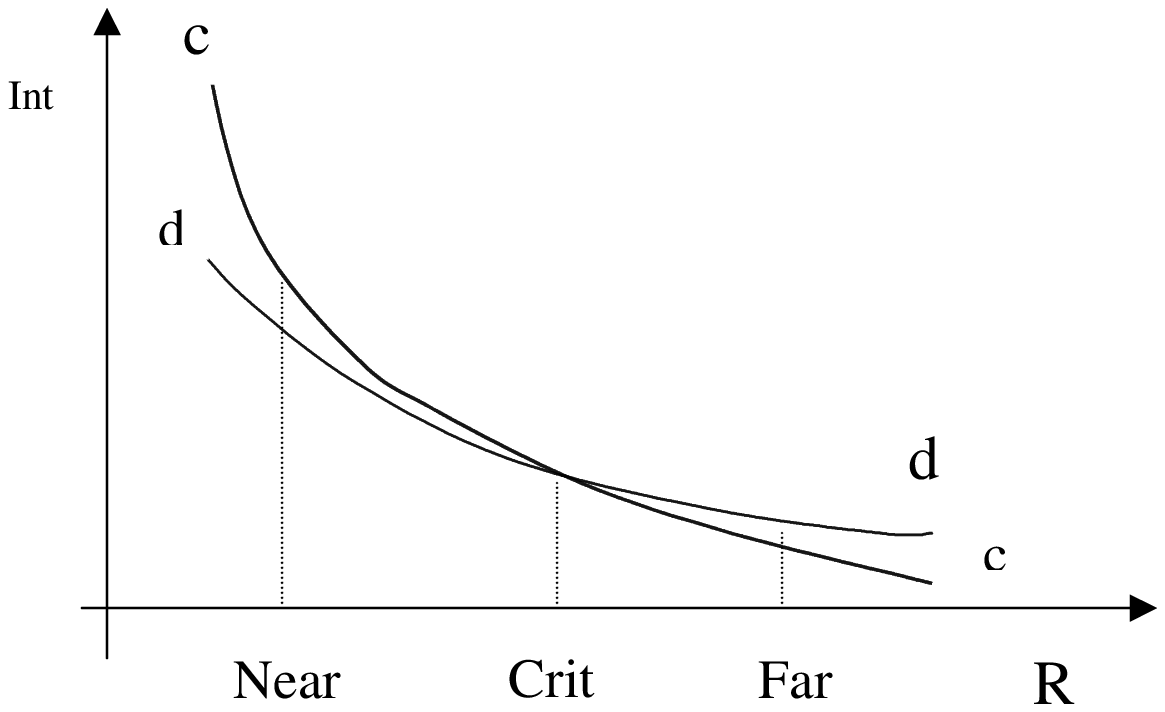}
\end{figure}}
\end{minipage}\hfill
\vglue-8cm
\hglue-1.0cm
\begin{minipage}[]{7.0cm}
\begin{figure}
\hglue8.0cm
\parbox[t]{5.0cm}{\vglue-5.5cm
\caption{Two descriptions for a same interaction: continuous (cc) and discrete (dd). For convenience the discrete one is approximated by a smoothly 
continuous curve. The near and the far demarcating points delimit the thresholds of existing technology for resolving the two curves. The critical point, 
defined by the two-curve crossing is technologically independent and represents a fundamental scale for the interaction intensity in terms of an effective 
derivative of $q(\tau)$.}}
\end{figure}
\end{minipage}

\vglue-1cm
The two curves are just, respectively discrete and continuous,
representations of a given generic interaction. We are interested on their asymptotic regions where, in principle, discrepancies between them can be 
detected. An interaction where $\Delta\tau_{j}$ but not $\Delta q_{j}$ goes to zero with the distance $R$, diverges in the continuous description 
as $\frac{\Delta q_{j}}{\Delta\tau_{j}}$ goes to infinity  whereas it
remains finite in the discrete one. In the discrete description the interaction is always finite,  no matter how strong. It has been discussed in
\cite{hep-th/9610028,hep-th/0006214,gr-qc/9801040} for both the gravity and
the electromagnetic field. The inconsistencies of the continuous fields,
made explicit through divergences and causality violations, disappear with
the discreteness, with the existence of a non null lapse of time between
two consecutive interaction points, or in other words, with the recognition
that each interaction point is an isolated event. 

In the far asymptote, for an interaction with
\be
\Delta q_{j}\ge const>0,
\ee
${\dot q}$ in the continuous description goes to zero as $R$ (and therefore
$\Delta\tau_{j}$) goes to infinity whereas the discrete one tends
to a finite and constant value. It just becomes more and more intermittent but not necessarily
goes to zero.

At very large distances where $\Delta \tau_{j}$ becomes
detectable the field asymptotic limit should reveal its discrete
nature. Actually this possibility is spoiled, in the case of a
matter-polarizing field like the electromagnetic one, by the shielding
effect: The field is canceled before $\Delta \tau_{j}$ grows to the point
of detectability. This, of course, does not happen to gravity and so
effects of this expected discreteness must be observed but this discussion
will be deferred to Section VI.

Careful observation at both small and large distances for these cases should reveal that
the strength of the actual interaction $(\Delta q_{j})$, respectively,
grows and decreases at a smaller rate than the theoretical prediction from
a continuous interaction. When observed, in a context of continuous
interactions, these effects may require the use of regularization and
renormalization techniques or may give origin to various misleading
interpretations like the existence of new forms of fundamental continuous
interactions or of strange and yet to be observed form of matter, for
example.
Calculus (integration and differentiation) in a discrete-interaction context becomes useless for a rigorous description of fundamental physical processes. 
But in practice such a detailed strictly discrete calculus is not always necessary and in some cases may  not even be feasible.
 What effectively counts is the scale determined by $\Delta\tau_{j}$, the time interval between two consecutive interaction events, face the accuracy of the 
 measuring apparatus. The question is if $\Delta\tau_{j}$ is large enough to be detectable, or how accurate is the measuring apparatus used to detect it. 
 The density of interaction points on the world  line of a given point charge is proportional to the number of point charges with which it interacts. Let 
 one consider the most favorable case of a system made of just two point charges. As the argument is supposedly valid for all fundamental interactions one 
 can take the hydrogen atom in its ground state for consideration, treating the proton as if it were also a fundamental point particle. The order of scale 
 of $\Delta\tau_{j}$ for an electron in the ground state of a hydrogen atom is given then by the Bohr radius  divided by the speed of light
$$\Delta\tau_{j}\sim10^{-18}s$$ which corresponds to a number of
${\pi\over\alpha}\sim400$ interactions per period ($\alpha$ is the
fine-structure constant) or $\sim10^{10}$ interactions/cm. So, the electron
worldline is so densely packed with interaction events that one can, in
an effectively good description for most of the cases, replace the graph of
the Figure 2 by a continuously smooth curve. The validity of calculus in
physics is then fully reestablished in the interval between the two
demarcating points as a consequence of the limitations of the measuring
apparatus. The Wigner's questions\cite{Wigner} about the unexpected
effectiveness of mathematics in the physical description of the world is
recalled.  The answer lies on the huge number of point sources in
interaction (a sufficient condition), the large value of the speed of light
and the small (in a manly scale) size of atomic and subatomic systems,
which indirectly is a consequence of h, the Planck constant.


Even in these situations where $\Delta\tau_{j}$ may not be measurable, at least with the present technology, the discrete formalism is justified not for 
replacing the continuous one where it is best, which       is confirmed by high precision experiments \cite{Nakanish,Darmour} but mostly for defining and 
understanding its limitations. There are, besides this very generic justification, many instances of one-interaction-event phenomena, like the Compton 
effect, particle decay, radiation emission from bound-state systems, etc, where discrete interactions are the natural and the more appropriate approach. 
These are, of course, all examples of quantum phenomena, but primarily because quantum here implies discreteness.

\subsection{Discrete-continuous transition}

 It would be interesting to have a framework where this change from continuous to discrete interaction and vice-versa could be formally realized in a simple 
 and direct way.
One can deal with them considering the behaviour under a derivative operator of  ${\bar\theta}(\tau)$ which is the mathematical description of the 
interaction discreteness. Then one must require that, symbolically
\be
\l{dk}
\frac{\partial}{\partial\tau}{\bar{\theta}}(\tau-\tau_{i}):=\delta_{\tau\tau_{i}},
\ee
with $\delta_{\tau\tau_{i}}$ the Kronecker delta
\be
\delta_{\tau\tau_{i}}=\cases{1, &if $\tau=\tau_{i}$;\cr
    &\cr
    0, & if $\tau\ne\tau_{i}$,\cr}
\ee
with the meaning that at the points where the left-hand side of Eq. (\ref{dk}) is not null, which are the only relevant ones,  $\tau$ must be treated as a 
discrete variable and that the operator $\frac{\partial}{\partial\tau}$ must be seen as (or replaced by) just a sudden increment $\Delta$ and not as the 
limit of the quotient of two increments.

Then with such a convention one has from Eq. (\ref{dV}) that
\be
\l{nabla}
\nabla_{\nu}q(\tau){\Big |}_{f}=-f_{\nu}\sum_{i}q_{\tau_{i}}\{{\bar\theta}(\tau_{i+1}-\tau)\delta_{\tau\tau_{i}}-\delta_{\tau_{i+1}\tau}{\bar\theta}
(\tau-\tau_{i})\}:=-f_{\nu}{\dot q}(\tau),
\ee
which implies that ${\dot q}(\tau)$ is null when $z(\tau)$ is not a point of interaction on the charge world  line. For such an interaction point 
$\tau_{j}$ one has
\be
{\dot q}(\tau_{j})
=q_{\tau_{j}}{\bar\theta}(\tau_{j+1}-\tau_{j})-q_{\tau_{j-1}}{\bar\theta}(\tau_{j}-\tau_{j-1})=q_{\tau_{j}}-q_{\tau_{j-1}}
\ee
or, generically
\be
\l{ac}
{\dot q}(\tau)=\cases{\Delta q_{i}=q_{\tau_{i}}-q_{\tau_{i-1}} & for $\tau=\tau_{i}$;\cr
        &\cr
        0 & for $\tau\ne\tau_{i}$,\cr}
\ee
and, from the middle term of Eq. (\ref{nabla})
\be
\nabla^{f}_{\sigma}\nabla^{f}_{\nu}q(\tau)=-2f_{\sigma}f_{\nu}\sum_{i}q_{\tau_{i}}\delta_{\tau\tau_{1}}\delta_{\tau_{i+1}\tau}=0.
\ee
 In Eq. (\ref{nabla}) $i$ labels the vertices and only these points on the world
 line contribute. That is why one has to define  Eq. (\ref{dk}). In a limit where a
 summation over $i$ may be approximated by a time integration the Kronecker delta may
 be replaced by a Dirac delta function and then one may have Eq. (\ref{dasdd}) as a
 good operational approximation to Eq. (\ref{ac}).\\
Therefore we understand Eqs.(\ref{AfV},\ref{dAf}) as meaning, respectively
\be
\l{dAj}
\phi_{f}(x)=q(\tau){\Big |}_{f}=\cases{q_{\tau_{j+1}}{\Big |}_{f} &if $\tau_{j}<\tau_{ret}<\tau_{j+1}$\cr
        &\cr
        &\cr
         \frac{q_{\tau_{j+1}}+q_{\tau_{j}}}{2}{\Big |}_{f}& if $\tau_{ret}=\tau_{j}$\cr}
\ee
and
\be
\l{dAjj}
\nabla_{\mu}\phi_{f}(x)=-f_{\mu}\Delta q(\tau){\Big |}_{f}=\cases{-f_{\mu}(q_{\tau_{j+1}}-q_{\tau_{j}}){\Big |}_{f} &if $\tau_{ret}=\tau_{j}$\cr
        &\cr
         0& if $\tau_{ret}\not=\tau_{j}$\cr}
\ee The field $\phi_{f}(x,\tau)$ is just like an instantaneous picture of
its source at its retarded time; a travelling picture. If $z(\tau_{ret})$
is not a point of change in the source's state, $\phi_{f}(x)$ is not
endowed with a physical  meaning as its energy tensor is null. A physical
discrete field always corresponds to a sudden change in its source's state
at its retarded time. If there is no change the field is not real, in the
sense of having zero energy and zero momentum. Having no physical attribute
it corresponds to a pure ``gauge field" of the continuous formalism.

\section{Scalar field and general relativity}

It takes an external agent to cause a change $\Delta q$ on the charge $q$ of a
scalar source; a positive $\Delta q$ means that a scalar field $\phi_{f}(x,\tau)$
has been, say, absorbed whereas a negative one means then an emission. Therefore, a
discrete scalar field carries itself a charge $\Delta q$ and can, consequently,
interact with other charge carriers and be a source or a sink for other discrete
scalar fields. It carries a bit of its very source, a scalar charge; it is an
abelian charged field.
 On the other hand a new look at equations (\ref{psf}) and (\ref{ac}) reveals that
 $(\Delta q_{j})^2$ describes the energy-momentum content of the field. So, the
 source of a discrete scalar field is any physical object endowed with energy which
 corresponds then to the scalar charge.  Energy, of course, is a component of a four-vector and not a Lorents scalar. Its four-vector character comes from the $f^{\mu}$ factor in Eq. (\ref{psf}): the energy of $\phi_{f}(x,\tau)$ is the fourth component of the current of its squared scalar charge. The scalar charge conservation is therefore assured  by and reduced to the conservation of energy and momentum given by Eq. (\ref{emc}).  Considering the relativistic mass-energy relation this implies that the discrete scalar field satisfies the Principle of Equivalence and that all physical objects interact with the scalar field through its energy-tensor.  This is a form of the Principle of Universality of gravitational interaction, introduced by Moshinski\cite{Moshinsky}. So, $\phi_{f}(x,\tau)$ must necessarily be connected to the gravitational field. Having necessarily energy for source implies on an important consequence of uniqueness, of excluding the existence of any other distinct fundamental discrete scalar field as it must necessarily be taken as the gravitational field\footnote{There would be no point on assuming that a same charge could be the source of two or more distinct fields with the same characteristics}. Moreover, as energy is not a scalar, the symmetry between discrete fields and  sources, both taken as fundamental fields, implies also that there should be no fundamental scalar source representing an elementary field; it must be a scalar function of a non-scalar fundamental field, like the trace of an energy tensor, for example. This lets then explicit a known symmetry of nature: the four fundamental interactions are described by gauge fields having vector currents for sources ($j=qv,$ as they are pointlike sources), including gravity since the energy tensor is just a current of its charge, the four-vector momentum. So, this symmetry is not broken with gravity being a second-rank tensor field.\\
This possible physical interpretation is compatible with the General Theory of Relativity, according to the work done in the references  \cite{gr-qc/9801040,gr-qc/9903071}, where a discrete gravitational field defined by
\be
\l{gmn}
g_{\mu\nu}^{f}(x)=\eta_{\mu\nu}-\chi
f_{\mu}f_{\nu}\phi_{f}(x,\tau),
\ee
as a point deformation in a Minkowski spacetime, propagating on a null direction $f$, upon an integration on $f$, in the sense of Eq. (\ref{dAf}), reproduces the standard continuous solutions. That gravity be either totally \cite{inMTW} or partially \cite{Fierz,Darmour} described by a scalar (continuous) field is an old idea\cite{FierzePauli,Jordan,Brans-Dicke}, but Eq. (\ref{gmn}) implies on regarding gravity as being ultimately described by a discrete scalar field in a metric theory. With the metric in this form the Einstein's field equations
\be
\l{Eeq}
R_{\mu\nu}-{1\over2}g_{\mu\nu}R=\chi T_{\mu\nu}
\ee
is reduced \cite{gr-qc/9801040} to
\be
\l{rEeq}
f_{\mu}f_{\nu}\eta^{\alpha\beta}\nabla_{\alpha}\nabla_{\beta}\phi_{f}(x,\tau)=\chi T_{\mu\nu},
\ee
as the gauge condition used in \cite{gr-qc/9801040}
\be
f^{\mu}\nabla_{\mu}\phi_{f}(x,\tau)=0
\ee
becomes an identity after Eq. (\ref{dAf}), as $f^2=0$.

Inherent to discrete fields, irrespective of their tensor or spinor
character, is the  implicit  conservation of their sources as a consequence
of their (discrete fields) very definition\footnote{Schematically:
$j^{\mu}=qv^{\mu}$ $\Rightarrow$
$\nabla_{\mu}j^{\mu}=-qa^{\mu}f_{\mu}\equiv 0$ as $a.f\equiv 0,$ according
to Eq. (22) of \cite{hep-th/0006237}.}. This is discussed in Section V of
paper I. So, whereas $T^{\mu\nu};_{\mu}=0$  is assured by the symmetry of
the Einstein tensor on the left-hand side of Eq. (\ref{Eeq}), in Eq.
(\ref{rEeq}) it is just a consequence (see Eq.(\ref{emc})) of Eq.
(\ref{dAf}). This symmetry of the Einstein tensor is in this way similar to
the one of the Maxwell tensor that assures charge conservation in the
standard continuous-field formalism but that is a consequence of extended
causality (discrete-field definition) and Lorentz symmetry
\cite{hep-th/9911233} in a discrete-field approach.

The Eq. (\ref{gmn}) reminds an old derivation \cite{Feynman} of the field equations of general relativity by consistent re-iteration of
\be
\l{reit}
g_{\mu\nu}(x)=\eta_{\mu\nu}+\chi h(x)_{\mu\nu},
\ee
as solution from a gauge invariant wave equation for the field $g_{\mu\nu}(x)$ in a Minkowski spacetime. The non-linearity of the Einstein's equations 
comes from contribution to $g_{\mu\nu}(x)$ from all terms of higher orders in $h_{\mu\nu}$. Therefore, the 
results obtained in the reference \cite{gr-qc/9801040} imply that if $h_{\mu\nu}$ is ultimately a discrete scalar field 
$$h_{\mu\nu}=f_{\mu}f_{\nu}\phi_{f}(x,\tau),$$ there is no higher order contribution essentially because $f^2=0$. A discrete field has no self-interaction, a 
consequence of its definition (\ref{df}) and that is explicitly exhibited in its Green's function (\ref{pr9}). Discrete fields are solutions from linear  
equations. Whereas this is true for $g^{f}_{\mu\nu}$ of Eq. (\ref{gmn}) it is not for its $f$-averaged $g_{\mu\nu}$ of Eq. (\ref{reit}). The non-linearity of 
general relativity appears here then as a consequence of the averaging process of Eq. (\ref{s}) that effectively smears the discrete field over the 
lightcone, erasing all the information contained in $f$. The interested reader is addressed to the references \cite{gr-qc/9801040} and \cite{gr-qc/9903071}.

On the other hand  the energy tensor in Eq. (\ref{rEeq}) must be traceless, also a consequence of $f^2=0$. This reminds an old known problem in standard 
field theory that comes when a massless theory is taken as the $(m\rightarrow0)-$limit of a massive-field theory 
\cite{BoulwareeDeser,Deser,Van-Dam,Feldman,Fierz}, but for a discrete field, in contradistinction, a traceless tensor does not necessarily mean a massless 
source \cite{hep-th/0006237}.
The wave equation (\ref{rEeq}) must be preceded by some careful qualifications, however. A discrete field requires a discrete source. The source in 
Eq. (\ref{rEeq}) must be treated as a discrete set of point sources $T^{f}_{\mu\nu}(x,\tau)$ for which $f^{\mu}T^{f}_{\mu\nu}(x,\tau)=0$. This implies that 
there is no exterior solution for a discrete gravitational field, only vacuum solutions. Any interior continuous solution must be seen then as an 
approximation for a densely packed set of point sources. From the discrete vacuum solution of Eq. (\ref{rEeq}) one can, in principle, with an integration 
over its $f$-parameters, obtain any continuous vacuum solution of
an imposed chosen symmetry\footnote{From the superposition of the discrete fields of a spherical distribution of massless dust one retrieves the Vaydia 
metric\cite{unpublished}.}\cite{gr-qc/9801040}. This justifies, up to a certain point, not regarding the right-hand side of Eq. (\ref{gmn}) as just the 
first two terms of a series of possible contributions from higher rank tensors.
 Even for a massive point-source, however, being itself a discrete field, $T^{f}_{\mu\nu}$ cannot be expressed in terms of its mass and of its actual 
 four-velocity $v$. A traceless $T^f_{\mu\nu}(x,\tau)$ with $f^{\mu}T^{f}_{\mu\nu}(x,\tau)=0$ does not necessarily represent a massless source nor $f$ 
 represents its four-velocity, as discussed in Section V of paper I.

The geometrical description of gravity as the curvature of a
pseudo-Riemannian spacetime has its validity, in the range of the ratio
parameter (\ref{ratio}) limited by the two demarcating points of the Figure
4, always assured as an absolutely good approximation due to the high
density of interaction points in any real measurement, as discussed in the
previous section.

\section{Theory and observations: possible links}
In this section we want to make some brief comments on some possible
theoretical and observational evidences of direct consequences of
interaction discreteness, particularly in gravity.  The comparison between
the discrete and the continuous description  of an interaction leads to the
existence of the critical point and of experimental thresholds (near and
far) for resolving the interactions, as shown in the Figure 4. The interior
segment, between these two values, defines the domain of validity of the
continuous-interaction approximation where the polygonal worldline of the
sources are so densely packed of interaction points that they can be
effectively replaced by smoothly continuous curves and the concept of
acceleration and of spacetime curvature at a point on the worldline make
sense.  We are not proposing, it is worth emphasizing, the replacement of
general relativity in its domain of validity by a discrete scalar field
theory of gravity, and similar statements should be assumed for other field
theories. The point is that in this domain, i.e. for $\frac{\Delta
q_{j}}{\Delta\tau_{j}}$ between the two demarcating points, it cannot make,
by definition, any experimentally detectable difference. Considering the
small strength of its coupling the gravitational interaction is irrelevant
for physical systems involving relatively few fundamental elements. Even a
gravitational Aharanov-Bohm-like experiment \cite{Sakurainaoda!} would
require the gravitational field of a macroscopically large object, like the
Earth. The sufficient condition for a high density of interaction points is
assured and justifies continuous descriptions of gravity, of which general
relativity seems to be the best proposal \cite{Darmour}. Moreover  the
undectability of discrete gravity in this region is tantamount to the
unobservability of the Minkowski spacetime. At this level  of approximation
the Minkowski spacetime becomes the local tangent space of an effective
curved space-time and $f$ a generator of the local hypercone in its tangent
space. This would lead to full general relativity in accordance to a
general uniqueness result \cite{MTW,Visser} that any metric theory with
field equations linear in second derivatives of the metric, without
higher-order derivatives in the field equations, satisfying the Newtonian
limit for weak fields and without any prior geometry must be exactly
Einstein gravity itself. This reminds  again the already mentioned
\cite{Feynman} derivation of general relativity from flat spacetime but now
with the distinctive aspect that the effective Riemannian spacetime comes
not from a consistency requirement but as an approximation validated by the
limitation of our experimental capacity, which can always be improved, be
placed on more stringent limits, but never be totally eliminated.

On the other hand, outside this region, i.e. below or above the thresholds,
the discrepancies between a discrete and a continuous interaction cannot be
overlooked. This casts doubts on the results about asymptotic fields and
their singularities of any continuous-field theory. By the way, considering
that the discrete field is weaker than the continuous one  in the origin
neighborhoods we can suggest or expect that the discrete field may give an explanation
to inflation or at least alleviate its need in cosmological theories.

\subsection{Discrete Newtonian potentials}

The evolution of a system through a sequence of $n$ discrete interactions
is described by a series involving combinatorials of $n$, i.e.
$n(n-1)(n-2)\dots$ This is a natural consequence of discrete interactions:
power series replacing continuous functions obtained from integrations of
differential equations. The evolution of any system is given in terms of
power series. A continuous interaction, irrespective of its duration, would
always be equivalent to an infinite n. This is the meaning of a
conservative potential  and this is why a continuous interaction invariably
has problems with infinities. Just for the sake of illustrating this very
important point let us, anticipating some results\footnote{This will be
presented with details elsewhere. Its anticipation here is just for the
sake of illuminating the arguments.}, consider the much simpler case of a
radial motion with a non-relativistic axially symmetric interaction (a
logarithmic effective potential, an effective
inversely-proportional-to-the-distance acceleration). This symmetry implies
that the change in speed at each interaction is a (very small) constant
$\Delta$. For initial conditions taken, right after an interaction event,
as $$r(t_{0})= r_{0};$$ $$v(t_{o})=v_{0},$$ the next interaction will occur
at $$t_{1}= t_{0}+\Delta t_{0}=t_{0}+\alpha r_{0},$$ where $\alpha$ is also
a very small constant, and $$v(t_{1})=v_{1}=v_{0}-\Delta;$$ $$r(t_{1})=
r_{0}+v_{0}\alpha r_{0}=(1+\alpha v_{0})r_{0},$$ as there is free
propagation between any two consecutive interactions.
 Therefore, for the $n^{th}$ interaction
\be
\l{rn}
r_{n}=r_{n-1}+v_{n-1}\Delta t_{n-1}=(1+\alpha v_{n-1})r_{n-1}=r_{0}\amalg_{i=0}^{n-1}(1+\alpha v_{i}),
\ee
 with
\be
\l{vn}
v_{i}=v_{0}-i\Delta.
\ee
Then, from Eq. (\ref{rn}),
\be
\frac{r_{n}}{r_{0}}= 1+\alpha\sum_{i_{1}=0}^{n-1}v_{i_{1}}+\alpha^2\sum_{i_{1}=0}^{n-1}\sum_{i_{2}=i_{1}+1}^{n-1}v_{i_{1}}v_{i_{2}}\dots+\alpha^{n-1}
(\sum_{i_{1}=0}^{n-1}\sum_{i_{2}=i_{1}+1}^{n-1}\dots \sum_{i_{n-1}=i_{n-2}+1}^{n-1})v_{i_{1}}v_{i_{2}}\dots v_{i_{n-1}},
\ee
a finite series that with the use of Eq. (\ref{vn}) exhibits the following structure
$$
\frac{r_{n}}{r_{0}}= 1+\alpha({n\choose 1}v_{0}-{n\choose 2}\Delta)+\alpha^{2}[v_{0}^2{n\choose 2}-v_{0}\Delta(-3{n\choose 3}+2{n\choose 2}{n\choose 1}-2{n\choose 2})+$$
\be
+\Delta^2(-3{n\choose 4}+{n\choose 2}{n\choose 2}-4{n\choose 3}-{n\choose 2})]+{\cal{O}}(\alpha^3).
\ee
If $n>>1$, by considering just the largest contribution from each term in this finite series we have
\be
\frac{r_{n}}{r_{0}}= 1+\alpha n(v_{0}-\frac{n\Delta}{2})+\frac{\alpha^2n^2}{2}(v_{0}-\frac{n\Delta}{2})^2+{\cal{O}}(\alpha^3),
\ee
or
\be
\l{sn}
\frac{r_{n}}{r_{0}}=\sum_{k=0}^{n-1}\frac{1}{k!}[\alpha n(v_{0}-\frac{n\Delta}{2})]^k.
\ee
From Eq. (\ref{vn}), we have $$n=\frac{v_{0}-v_{n}}{\Delta},$$ which in Eq. (\ref{sn}) produces
\be
\frac{r_{n}}{r_{0}}=\sum_{k=0}^{n-1}\frac{1}{k!}[\frac{\alpha}{\Delta}\frac{(v_{0}^2-v_{n}^2)}{2}]^k.
\ee
The bigger is $n$ the better this finite series can be approximated by an exponential
\be
\frac{r_{n}}{r_{0}}\approx exp(\frac{\alpha}{\Delta}\frac{(v_{0}^2-v_{n}^2)}{2}),
\ee
which can be re-written as
\be
\frac{v_{0}^2}{2}+\frac{\Delta}{\alpha}\ln{r_{0}}\approx\frac{v_{n}^2}{2}+\frac{\Delta}{\alpha}\ln{r_{n}}=const.
\ee
This is energy conservation with an effective potential energy given by
\be
\l{Up} U(r)= \frac{\Delta}{\alpha}\ln{r}. \ee Then, for the gravitational
interaction we identify the constants as $$GM=\frac{\Delta}{\alpha},$$
where $M$ is the central mass. An infinite $n$ would make the right-hand
side of Eq. (\ref{Up}) to be an exact expression (in the corresponding
classical, non-relativistic limit) for the effective potential energy but as $n$ may at most be a huge but finite number
this represents just the sum of the largest contribution from each term in
this series. In other words, the right-hand side of Eq.
(\ref{Up}) is just an effective expression with a large but limited
domain of validity due to neglecting the smaller terms in the
combinatorials.  So, remarkable here is not only the appearing of the
Newtonian potential as an effective field but also its asymptotic
character: Energy is conserved at each interaction but the exact
mathematical expression of the potential energy is given by Eq.(\ref{Up})
only after an infinite number of interactions.

\subsection{The essential question}

The essential question that is posed now is which is the true nature of
fundamental interactions: Continuous or discrete?  This must be an experimentally based decision
but there are some arguments favoring\footnote{The reasons have been detailed on the references 
\cite{hep-th/0006237,hep-th/9610028,hep-th/0006214,gr-qc/9801040,hep-th/9911233,hep-th/9610145,gr-qc/9903071}. 
Parts of the old ones may have been superseded by the more recent ones.} the discrete case:

Continuous interactions are plagued by infinities and causality problems. They are inherent to the
continuous hypothesis. The discrete interaction is free of them and can profitably reproduce the entire
continuous formalism in terms of effective continuous interactions. The continuous case is contained in the discrete one. 
The immediate profits are the many
ad hoc features of continuous fields but that are natural consequences of either a discrete field or from
the discrete-to-continuous passage. The following subsection considers further implications of interaction discreteness.

\subsection{Boltzmann and Tsallis Statistics}

With discrete interactions we, rigorously, do not have differential equations nor integrations. The evolution of any system is done through sudden and 
discrete
finite differences that are just superimposed. Between two consecutive interaction points every point like component just moves freely on 
straight lines. All exact physical statements are expressed as finite power series involving those combinatorials. This is a general statement in the 
sense that
any physical system, even a macroscopic one, composed by an immense number of point like fundamental elements has its states, its conservation laws,
its evolution, its statistical distributions, etc. described in terms of power functions. This is so because there are no exact smoothly continuous 
solutions but segments of straight lines or as an idealized limit which should be attainable only after an infinite number of steps. An infinite number does not exist, and infinity is
just an idealized concept of a limit, of an unreachable boundary. Being so, the world is surprisingly simpler and our standard vision of it is richer of
such idealized, unreachable concepts than we had previously conceded. A whole paraphernalia of mathematical tools, so useful in physics - differential
equations, integrations, differential geometry, topology, just for citing a few - and so many familiar and daily used mathematical functions like sine,
exponentials, harmonic and coulombian potentials, circles, ellipses, etc., etc.,  do not belong to the realm of the physical world; they are just
unreachable, idealized limiting boundaries as much as an ideal gas and a macroscopic reversible process.

This supports the generalized one-parameter power function definition of entropy
introduced in 1988 by Tsallis [35], which  provides a power-law distribution of probabilities. 
The number of its application to the most diverse systems has, since then, steadily and
rapidly increased [36].  
 It is
reduced to Boltzmann statistics when its parameter is equal to unity. This
parameter is then a measure of how close the system is from its idealized
asymptotic state, that rigorously, is reachable only after an infinite
number of interactions. It is a proper statistics for a world made 
of discretely interacting point like objects. The Boltzmann statistics, as
any mathematical formulation  for physics, based on continuous
interactions, is displaced, according to this viewpoint, to these idealized
boundaries. But, of course, an immense $n$, in most cases, is an excellent
approximation to infinity.
The extensive applicability of Tsallis statistics on the most diverse real systems may be an indication of the true nature of 
the world, if continuous or absolutely discrete.
\subsection{Possible experimental evidences}

On the observational side we note that for the asymptotic region above the
critical point the continuous asymptotically null fields are replaced by
discrete interactions that become more and more intermittent with the
distance, but do not necessarily go to zero. This may be detectable for
the gravitational field as it does not have shielding effects although it
requires huge masses for detecting very weak gravitational fields and huge
distances for producing a detectable $\Delta\tau_{j}$; both conditions
found at and above galactic scales. Therefore, a right place for  checking
for signs  of discreteness may be the rotational dynamics of galaxies which
is essentially given by
\be
\l{dyngalax}
\frac{GMm}{R^2}=\frac{mv^2}{R},
\ee
so that the orbital velocities of galaxies would be expected to be inversely proportional to
the square root of the radial distance from the central mass. But both
sides of this equation are heavily dependent on the assumption of a continuous
interaction. The Newtonian field is a consequence, in a discrete interaction
context, of a large frequency of interaction points and, therefore, of a small
$\Delta\tau_{j}$. This is explicitly shown in
\cite{hep-th/0006237,gr-qc/9801040,hep-th/9610145}.
The centripetal force is an expression of inertia in a circular motion but for
discrete interaction the circle is replaced by a polygon as the body freely
moves on a straight line between two consecutive interaction events. Let us consider a
polygon circumscribed on a circle of radius R. Then
\be
\frac{v}{c}=\frac{\Delta x}{2R}\sim\frac{2\pi R}{n2R}=\frac{\pi}{n}, \ee
where $c$ is the speed of light, $n$ is the (enormous) number of
interaction events (the number of vertices) that, may
depend on $v$, but not on $R$. Then the
orbital velocity becomes independent\footnote{Another way of seeing it is that both $\Delta x_{j}$ and
$\Delta\tau_{j}$ are proportional to $R$.} of $R$ after the critical point.

So, flat rotation curve is something very natural in a discrete-field
context!  It is therefore a real possibility that the critical point for
gravity has already been detected in the flat rotation curves of galaxies
\cite{rotation curves of galaxies}. The flatness feature of a rotation
curve of a galaxy, as remarked by Milgrom \cite{Milgrom}, is determined not
by its central mass $M$ alone nor just by the distance $R$  but by the
acceleration which is equivalent to the ratio-parameter (\ref{ratio}) as
$\Delta q_{j}$  for gravity corresponds to a change of speed. Therefore the
existence of the critical point in the continuous/discrete physical
description justifies the introduction of a new fundamental scale for the
interaction strength in terms of an effective acceleration. This may put
Milgrom's MOND \cite{Milgrom} on a more sound physical basis.
 The actually prevailing wisdom that a flat rotation curves is the (ad hoc) indication of some strange, ubiquitous but still to be detected cold dark matter
 is not free of problems and is far from being unanimous\cite{Milgrom,cdm,Evans,Mannheim,Bekenstein,Nucamendi}.

Another possible evidence of discrepancy that must be considered is the apparent anomalous, weak, long-range acceleration observed in the Pioneer 10/11,
Galileu, and Ulysses data \cite{Pioneer}. Due to their spin-stabilization and to the great distance (30 t0 67 AU) from the Sun the spacecrafts are excellent
for dynamical astronomy studies as they permit precise acceleration estimation to the level of $10^{-10}cm/s^{2}.$ The detected anomalous acceleration comes
from the second largest contribution from those mentioned $n$-combinatorials. Eq. (\ref{rn}) is, of course, not valid for circular motion, and so there is 
no second largest contributions and, therefore, no Pioneer effect on planetary orbits \cite{JDA}. Both cases, the rotation curves and the spacecraft dynamics, in the 
context of
discrete interactions, will be discussed with details elsewhere.

\section{Conclusions}

The thesis that fundamental interactions are discrete is being developed. If this is the case there is no really compelling reason for excluding gravity
from such a unifying idea.
The knowledge of a supposedly true discrete character of all fundamental interactions is a permanent reminder of the limits of a continuous approximate 
description. The idea of an essential continuity of any physical interaction allows unlimited speculations that will always go beyond any level of possible 
experimental verifications which brings then the risk of not being able of distinguishing  the reign of possibly experimentally-grounded scientific 
research  from plain philosophical speculation or even just fiction. Regardless the possibility that some of its consequences have already been 
experimentally detected, a discrete gravitational interaction, even in the range where it is not experimentally detectable, still for a long time to come, 
may just make sense of existing theories for delimiting their domain of validity  as it has historically happened with all new discreteness introduced in 
the past, like the ideas of molecules, atomic transitions, and quarks, for example.


\begin{thebibliography}{10}
\bibitem {hep-th/0006237}M.M. de Souza, {\it Conformally symmetric massive discrete fields}, submitted to publication. hep-th/0006237.
\bibitem{Polchinski}J. Polchinski, {\it String theory. Vol. I}, Cambridge Univ. Press, Cambridge,(1998).
\bibitem {hep-th/9610028} M. M. de Souza, J. of Phys. A: Math. Gen. 30 (1997)6565-6585. hep-th/9610028
\bibitem{Wigner} E. P. Wigner, {\it The Unreasonable Effectiveness of Mathematics in Natural Sciences} reprinted in
{\it The Collected Works of Eugene Paul Wigner}, Vol VI, part B, J. Mehra ed., Springer Verlag, Berlin(1995); R.Jackiw {\it The Unreasonable Effectiveness of Quantum Field Theory} hep-th/9602122.
\bibitem{hep-th/0006214}M. M. de Souza, {\it Dynamics and causality constraints}. hep-th/0006214.
\bibitem{Jackson} D. Jackson {\it Classical Electrodynamics},2nd ed., chaps. 14 and 17,
John Wiley {\&} Sons, New York, NY(1975).
\bibitem {gr-qc/9801040} M. M. de Souza, Robson N. Silveira,  Class. \& and Quantum Gravity, vol 16, 619(1999). gr-qc/9801040.
\bibitem{inMTW}G. Nordstrom, Ann. Phys. (Germ.)   42, 533(1913); see Section 17.6 of reference \cite{MTW}.
\bibitem{MTW}C.W. Misner, K.S. Thorne, J.H. Wheeler, {\it Gravitation}, Freeman,S. Francisco(1973).
\bibitem{FierzePauli}M. Fierz, W. Pauli, Proc. Roy. Soc., 173,211(1939).
\bibitem{Jordan}P.Jordan, Nature 164, 637(1949).
\bibitem{Brans-Dicke}C. Brans, R.H. Dicke, Phys. Rev. 124, 925(1961),{\it ibid.}125,2194(1961).
\bibitem{hep-th/9911233} M.M. de Souza,{\it {Discrete gauge fields}} hep-th/9911233
\bibitem {hep-th/9610145} M. M. de Souza, {\it Classical Fields and the Quantum concept.}hep-th/9610145.
\bibitem{BoulwareeDeser}D.G. Boulware and S. Deser, Phys. Rev. D6,3368(1972).
\bibitem{Deser}S. Deser, J. Gen. Rel. Grav.,1,9(1970).
\bibitem{Van-Dam}H. Van Dam, M. Veltman, Nucl. Phys.,B22,397(1970).
\bibitem{Feldman}G. Feldman,{\it Classical Electromagnetic and Gravitational Field Theories as Limits of Massive Quantum Theories} in
{\it The Uncertainty Principle and Foundations of Quantum Mechanics; a fifty years survey},W.C. Price and S.S. Chissick (eds.)- John Wiley
\& Sons, London(1977).
\bibitem{Fierz}M. Fierz, Helv. Acta 29, 128(1956).
\bibitem{Darmour}C.M.Will, {\it Theory and Experiment in Gravitational Physics}, 2nd. ed. Cambridge University Press, Cambridge, 1993; T.Darmour,
{\it Experimental tests of Relativistic Gravity}, gr-qc/9904057.
\bibitem{Sakurainaoda!}R. Colella, A.W. Overhauser, S.A Werner. Phys. Rev. Lett. 34, 14722(1975). See also the references in G.Z. Adunas, E. Rodriguez-Milla, D.V. Ahluwalia, {\it Probing quantum violations of the equivalent principle} gr-qc/0060022.
\bibitem{unpublished}M. M. de Souza, unpublished.
\bibitem{Nakanish} T. Kinoshita and D.R. Yennie in {\it Quantum Electrodynamics} T. Kinoshita (ed.), World Scientific, Singapore(1990).
\bibitem{Feynman}W. Wyss,Helv. Phys. Acta,38(65)469;S. Deser, Gen.Rel.Grav,1(1970) 9.
\bibitem{Moshinsky}Moshinsky M., Phys. Rev. v. 80,514(1950); R.P.Feynman, F.B.Morinigo and W.G. Wagner,{\it Feynman Lectures on
Gravitation},B.Hatfield(ed),Addison-Wesley,Reading(1995).
\bibitem{gr-qc/9903071}M. M. de Souza, Robson N. Silveira, {\it Gauge vs Coulomb and the cosmological mass-deficit problem}, gr-qc/9903071.
\bibitem{Visser}M. Visser {\it Mass for the graviton}, gr-qc/9705051
\bibitem{rotation curves of galaxies}Zwicky, F., Helv. Phys. Acta 6, 110(1933).
\bibitem{Milgrom}M. Milgrom, {\it The modified dynamics-a status review}. astro-ph/9810302.
\bibitem{cdm}J. A. Sellwood, A. Kosowsky, {\it Does Dark Matter Exist?} in Gas \& Galaxy Evolution, ASP Conference Series, Vol. , 2000. J. E. Hibbard, M. P.
Rupen, J. H. van Gorkom, eds. astro-ph/0009074.
\bibitem{Evans}N. W. Evans, {\it No need for dark matter in galaxies?}. astro-ph/0102082.
\bibitem{Mannheim} P. D. Mannheim, {\it Open questions in classical gravity}. gr-qc/9306025.
\bibitem{Nucamendi} U. Nucamendi, M. Salgado, D. Sudarsky, {\it An alternative approach to the galactic dark matter problem}. gr-qc/0011049.
\bibitem{Bekenstein}J. D. Bekenstein,{\it New gravitational theorie as alternatives to dark matter} in {\it Proceedings of the Sixth Marcel Grossmann 
Meeting on General Relativity}, Part A, ed. H. Sato and T. Nakamura, World Scientific, Singapore,(1992), p. 905.
\bibitem{Pioneer}J.D. Anderson, P. A. Laing, E. L. Lau, A. S. Liu, M. M. Nieto, S. G. Turyshev, Phys. Rev. Lett. 81, 2858(1998). gr-qc/9808081;
{\it The apparent anomalous, Weak, Long-Range Acceleration of Pioneer 10 and 11}, S.G. Turyshev, J. D. Anderson, P. A. Laing, E. L. Lau, A. S. Liu, M. M.
Nieto. gr-qc/9903024.
\bibitem{JDA} J. d. Anderson, J.K. Campbell, R. F. Jurgens, E. L. Lau, X. X. Newhall, M. A. Slade III, E. M> Standish Jr, {\it Recent developments in solar-system tests of general relativity} in {\it Proceedings of the Sixth Marcel Grossmann Meeting on General Relativity}, Part A, ed. H. Sato and T. Nakamura, World Scientific, Singapore,(1992), p. 353.
\bibitem{Tsallis} C. Tsallis, J. Stat. Phys. 52, 479(1988).
\bibitem{Tsallis homepage} See the Bibliography at http://tsallis.cat.cbpf.br/biblio.htm
\end{thebibliography}
\end{document}